# Observation of Nonequilibrium Carrier Distribution in Ge, Si and GaAs by terahertz-pump--terahertz-probe Measurements

János Hebling,[1,2] Matthias C. Hoffmann,[1] Harold Y. Hwang,[1] Ka-Lo Yeh,[1] and Keith A. Nelson[1]

[1]*Massachusetts Institute of Technology, Cambridge, MA 02139, USA*
[2]*Department of Experimental Physics, University of Pécs, 7624 Pécs, Hungary*

We compare the observed strong saturation of the free carrier absorption in n-type semiconductors at 300 K in the terahertz frequency range when single-cycle pulses with intensities up to 150 MW/cm$^2$ are used. In the case of germanium, a small increase of the absorption occurs at intermediate THz pulse energies. The recovery of the free carrier absorption was monitored by time-resolved THz-pump/THz-probe measurements. At short probe delay times, the frequency response of germanium cannot be fitted by the Drude model. We attribute these unique phenomena of Ge to dynamical overpopulation of the high mobility Γ conduction band valley.

The dynamics of hot carriers are influenced by interactions among carriers and between carriers and the host crystal lattice. Ultrafast optical and near infrared (near-IR) spectroscopy, including pump-probe measurements, have provided extensive information about carrier and quasi-particle dynamics in semiconductors[1]. These dynamics mediate the performance of many ultrafast electronic and optoelectronic devices.

In typical ultrafast pump-probe experiments, hot carriers are created by optical band-to-band transitions with photon energies exceeding the band gap. The dynamics of the carriers after the excitation can be followed by optical or THz probe pulses[2-4]. However, this method has limitations for investigation of carrier dynamics since both holes and electrons are created at the same time and carriers are usually created in a small region near the sample surface. Consequently, it is difficult to investigate single-component carriers in bulk materials by optical excitation.

Recently, the development of high-energy tabletop THz pulse sources[5,6] has enabled nonlinear transmission studies on semiconductors[7-9] as well as measurements with THz-pump and near-IR[10] or mid-IR[11] probe pulses. Using THz field strengths on the order of tens of kV/cm, ultrafast dynamics of impurities[7], excitons[10] and polarons[11] in GaAs and GaAs heterostructures have been revealed.

In this letter, we demonstrate the application of THz-pump/THz-probe measurements to the study of inter- and intra-valley dynamics of hot free electrons in three prototype semiconductors: Ge, Si and GaAs. The THz peak field strengths used in these experiments reached values of 150 kV/cm, comparable to those in nonlinear semiconductor devices such as Gunn-diodes and tunnel diodes. Electron heating by the THz pump pulse imparts on the order of 1 eV energy to the electrons, enabling a large fraction of the electrons to undergo intervalley scattering from the initial lowest energy conduction band valley into side valleys.

Because of the strong interaction of free carriers with THz radiation it is also possible to use single cycle THz pulses as a tool to monitor changes in carrier mobility and scattering rates[2-4]. Since the values for carrier mobility in side valleys are usually very different from the ones in the conduction band minimum, the change in the distribution of electrons between the different valleys can be followed by using a variably delayed THz probe pulse.

We used the same experimental setup as described in Refs. 12 and13. Near-single-cycle THz pulses with microjoule energies were generated in a LiNbO$_3$ (LN) crystal by optical rectification of 100 fs pulses at 800 nm with the tilted-pulse-front method[14-16]. For our experiments an amplified titanium sapphire laser system with 6 mJ pulse energy and 100 fs pulse duration at a repetition rate of 1 kHz was used. The optical beam was split using a 10:90 beamsplitter into two parts. The 10% part was passed through a chopper wheel and was used to generate the THz probe in the LN crystal. The 90% part was variably delayed and used to generate the THz pump pulse in the same LN crystal. The generated collinear THz pulses were collimated and focused by an off-axis parabolic mirror pair to a 1.2 mm diameter spot at the sample location where the maximum THz pulse energy was 2 µJ. The transmitted THz pulses were focused by another off-axis parabolic mirror pair into a ZnTe electro-optic (EO) sampling crystal where the THz field temporal profile was recorded by a variably delayed optical readout pulse[17] using a balanced detection amplifier. Chopping only the probe pulse and using lock-in detection ensured excellent suppression of the THz pump field. Fourier transformation and comparison of the THz probe field with and without the sample in place yielded THz absorption and dispersion spectra as a function of THz-pump/THz-probe delay. In addition to the time-resolved pump-probe measurements, nonlinear THz transmission measurements were conducted by using only the THz pump pulse and varying its intensity by rotating the first of two wiregrid polarizers inserted into its path.

Results of nonlinear transmission measurements[18] in a <111> oriented sample of n-type germanium are shown in Figure 1a. The 6-mm thick sample contained phosphorus impurities[19], identified by THz time domain spectroscopy (THz-TDS)[20] using low THz fields at 10 K, making the sample n-type with a free electron concentration of $n = (5\pm2) \cdot 10^{14}$ cm$^{-3}$ at room temperature. THz pulse energies and peak intensities reached 2 µJ and 150 MW/cm$^2$ respectively. At THz pulse energies larger than 0.2 µJ, we observe strong saturation of the free carrier absorption (Figure 1b). The absorption spectra at different field strengths, obtained by Fourier transformation of the THz field profiles, are shown in the inset of Figure 1a. At the maximum field strength, the absorption spectrum is strongly saturated over the frequency range between 0.2 and 1.8 THz. At low THz fields corresponding to pulse energies below 10 nJ, the shape of the absorption spectrum approaches the familiar Drude model shape observed in low-intensity THz-TDS measurements at room temperature (dotted line in Figure 1a). An unexpected behavior is apparent in the absorption spectra at intermediate pulse energy (100 nJ). In this case the absorption is still substantially saturated at low frequencies, but at frequencies above 1.0 THz the absorption is increased relative to the low intensity values. This behavior was not observed in similar intensity-dependent transmission measurements in GaAs or Si, although these materials also showed strong saturation of free-carrier absorption at high THz pulse energy.

Figure 1b shows spectrally integrated absorption data from intensity-dependent transmission measurements for Ge and n-type <100> oriented GaAs[12] (carrier concentration $N_c=8\times10^{15}$, thickness 0.4 mm) in which the incident THz pulse energy was varied by three orders of magnitude. At intermediate THz pulse energies (100 nJ) the spectrally integrated absorption in Ge is greater than in the unperturbed sample by up to 10% and up to 30% when integrated only in a frequency range between 1.0 and 1.3 THz. In contrast, the absorption in GaAs decreases monotonically with increasing THz pulse energy. This unusual behavior of Ge has not been observed in earlier absorption saturation measurements[21] conducted with narrowband 40 ns duration pulses at 606 GHz frequency.

To examine the dynamics of the electronic responses, THz-pump/THz-probe measurements were performed as described above. Figure 2 displays THz probe field profiles after transmission through the Ge sample for 1, 3 and 5 ps delay between the pump and the probe pulses. The corresponding absorption spectra are plotted in the inset. For short probe delays the absorption spectra of Ge (and not of GaAs and Si) are significantly different from a Drude-type free carrier spectrum. At low frequency the absorption shows a narrow peak, and at higher frequency the absorption is flat. For intermediate delays this flat part increases with increasing negative slope and the narrow peak becomes less pronounced. For long delays, the spectrum develops into a Drude-type form. The absorption measured at long delays is somewhat larger than the linear absorption because the probe pulse, though much weaker than the pump, still induces some nonlinear absorption as shown in Figure 1b. Contrary to Ge, GaAs and Si showed strong saturation at all frequencies measured, exhibited Drude-type spectra at all delay times, and recovered, but did not exceed, the original absorption strength over all frequencies at long times. The saturation and time-dependent recovery of the spectrally integrated absorption is shown in Figure 3 for Ge, GaAs and n-type <100> oriented Si (0.45 mm, $N_c$=5×10$^{14}$ cm$^{-3}$).

The free-carrier absorption coefficient in semiconductors at THz frequencies can be described approximately using the Drude model[22]:

$$\alpha_c(\omega) = \frac{\varepsilon_b \omega_p^2 \gamma}{nc(\omega^2 + \gamma^2)} = \frac{e^2 N_c \gamma}{\varepsilon_0 n c m^*(\omega^2 + \gamma^2)} = \frac{e N_c \gamma^2 \mu}{\varepsilon_0 n c (\omega^2 + \gamma^2)} \qquad (1)$$

where $\varepsilon_0$ is the permittivity of free space, $\varepsilon_b$ is the relative permittivity of the semiconductor in the absence of free carriers, $\omega_p$ is the plasma frequency screened by $\varepsilon_b$, $\gamma$ is the momentum relaxation rate, $n$ is the real part of the refractive index, $c$ is the speed of light in vacuum, $N_c$ is the free carrier concentration, $m^*$ and $e$ are the effective mass and electric charge of the free carrier and $\mu = e/(m^* \gamma)$ is the carrier mobility. The overall magnitude of $\alpha_c$ is proportional to the free carrier concentration and the mobility and inversely proportional to the effective mass. The observed saturation and subsequent recovery of the free-carrier absorption can be explained qualitatively in terms of electron scattering among different valleys. The inset of Figure 3a shows the simplified conduction band structure for Ge and GaAs. In the case of n-doped GaAs the Γ-valley has the lowest energy and is occupied initially before the arrival of the THz pulse. For Ge the L-valley has the lowest energy and it is initially populated. For Si there is no minimum at the Γ point and the X(Δ)-valley has the lowest energy.

A simple estimate based on the amount of absorbed THz energy and the number of carriers in the illuminated volume indicates that the average energy deposited per free carrier at the highest pump pulse intensity was 2.1 eV in GaAs and 0.9 eV in Ge, providing access to side valleys[22] at higher energies: the L (0.31 eV) and X (0.52 eV) valleys in GaAs and the Γ (0.14 eV) and X (0.19 eV) valleys in Ge. The energized electrons may be scattered into the side valleys during the THz pump pulse, because the intervalley scattering time is typically on the sub-picosecond timescale[23]. Since the mobility in the X valley of Ge is nearly five times smaller than in the initial L valley, scattering into the X valley decreases the free carrier absorption according to Eq. 1. In the L and X valleys of GaAs the effective mass is about three and six times larger respectively than in the initial valley[24]. Assuming comparable scattering rates in the different valleys, this means correspondingly lower mobilities in the L and X valleys. Therefore scattering into these side valleys results in a decrease of the THz absorption. Besides the mobility differences of the initial and side valleys, the nonparabolicity of the valleys can also result in a decrease in mobility and THz absorption with increasing carrier kinetic energy within a single valley. The nonparabolicity parameter reported for GaAs[24] indicates that adding $\approx 0.3$ eV energy to electrons in the lowest-energy valley reduces their THz absorption to about 1/3 of their initial value.

After the THz pump pulse leaves the sample, the electrons scatter among the initial and side valleys, the average electron energy decreases, and the absorption recovers. A three-level rate equation model for the populations of states in the initial and side valleys in GaAs (neglecting the high-lying X valley) was successfully demonstrated earlier[25] to simulate optical-pump/THz-probe experiments[2]. The inset to Figure 3b shows an extension of this model to four states to take into account the nonparabolicity of the initial valley. From the populations $n_i(t)$ obtained for the different states $i$ in the model, the overall absorption coefficient was calculated as $\alpha(t) \propto \sum_i \mu_i n_i(t)$, where $\mu_i$ is the mobility of state $i$. For GaAs, using this simple model and the previously reported intervalley scattering rates[25] $\gamma_{S3}=2$ ps$^{-1}$ and $\gamma_{3S}=20$ ps$^{-1}$ and intravalley relaxation rate[26] $\gamma_{3S}=20$ ps$^{-1}$ yielded a good but not perfect fit to the measured spectrally averaged time-dependent absorption, shown in Fig. 3b, with the best-fit parameter value $\gamma_{32}=9$ ps$^{-1}$. A rate equation model of this sort is inadequate to describe the dynamics of highly excited free carriers in detail. Monte Carlo simulations taking into account the full band structure[24] can provide more reliable results. In the absent of such calculations, we fit the spectrally integrated absorption recovery dynamics to single or multiple exponential decays in order to obtain approximate time scales for overall energy relaxation of the hot carriers in the different samples. For GaAs and Ge we obtained good fits (see Fig. 3) with single-exponential decay times $\tau_r = 1.9$ and 2.7 ps, respectively. For Si, a biexponential decay was necessary, with distinct time constants $\tau_{r1} = 0.8$ ps and $\tau_{r2} = 24$ ps. Since in Si there is only one rather high-lying (L) side valley that is accessible energetically, the fast decay may reflect the rate of relaxation out of this valley, and the slow decay the cooling of the electrons in the initial, nonparabolic [24] X valley.

Our most striking results are the increased (rather than decreased) Ge absorption at intermediate THz pulse energy (Figure 1b) and the non-Drude-type Ge transient absorption spectra observed for short probe delay times (Figure 2 inset). Compared to GaAs and Si, Ge is unique in that it has an intermediate-energy ($\Gamma$) valley whose mobility is higher rather than lower than that of the initially occupied (L) valley. Although the population of the $\Gamma$ valley is usually neglected for static cases[27] since the density of states (DOS) is 50 and 150 times lower than in the initial (L) and the highest lying (X) valleys respectively, its dynamical effects are strongly evident in our pump-probe and nonlinear transmission measurements. As we will explain, both unusual observations can be explained assuming significantly larger $\Gamma$ valley population than that expected from the DOS and a thermalized electron energy distribution among the valleys.

The non-Drude transient absorption spectra may arise from different Drude-type spectra associated with the different valleys, one component corresponding to a narrow absorption band from the Γ valley, and another component corresponding to an extremely broad absorption band from the other low-mobility valleys. As shown in the inset of Figure 2, a good fit to the observed absorption spectrum at 1 ps probe delay is possible assuming that 2.4% of the conduction band electrons are in the Γ valley and the momentum relaxation time is ~3 times longer (1.4 ps$^{-1}$) than in the L valley. The Γ valley population calculated from the DOS and thermalized electron energy distribution is only 0.5% (5 fold smaller) assuming parabolic bands, and about 0.02% (120 fold smaller) taking the nonparabolicity of the L valley into account.

We used a simple model to estimate the carrier absorption saturation as a function of THz pulse energy, assuming that the overall absorption was the sum of absorption by electrons in the different valleys and accounting for the nonparabolicity of the lowest valley only. The energy dependent distributions of electrons in the three valleys were calculated from the DOS $g_i(E) \propto (m_i^*)^{3/2}$ and the Fermi-Dirac distribution $F_i(E)$, and the absorption was calculated as

$$\alpha \propto \sum_i \int g_i(E) F_i(E) \mu_i(E) dE \qquad (2)$$

where in the lowest valley only, $m_i^*$ and $\mu_i$ were treated as functions of $E$.

For GaAs this simple model resulted in monotonically decreasing THz absorption as a function of THz pulse energy (or electron temperature), with a good fit to the measured data (Fig. 1b) assuming an electron temperature of 2200 K upon irradiation by a 2 μJ THz pump pulse. As indicated by the dotted curve in Fig. 1b, the result of a similar calculation for Ge also shows monotonically decreasing absorption with increasing THz pulse energy, without the maximum in absorption seen experimentally at intermediate THz pump pulse energies. However, a good fit is obtained with the same momentum relaxation rate found above, $\gamma = 1.4$ ps$^{-1}$, for the Γ valley, with the population in the Γ valley 50 times higher than its expected thermalized population, and with an electron temperature of 11,000 K at 2 μJ THz pump pulse energy, i.e. five times higher than the electron temperature in GaAs. Of course, comparable electron temperatures are expected for a given THz pulse energy in GaAs and Ge at the input faces of the samples. However, the electron temperature at the output face is eight times larger for Ge than for GaAs because far less of the THz pump pulse is absorbed in the Ge sample than in the GaAs sample. As a result, the average electron temperature throughout the sample is significantly larger in Ge than in GaAs. The electron temperature in Ge corresponds to average electron energy of 0.95 eV. When accounting for 20-30% energy dissipation to the lattice during the pump pulse[23], this corresponds to a total absorbed energy of 1.2 eV, comparing favorably to the estimate of 0.9 eV mentioned above.

Like the non-Drude Ge transient absorption spectrum, the increased Ge absorption at moderate THz pulse energies can be accounted for by the high mobility of the Γ valley, but only assuming a substantially higher transient population in that valley than determined by DOS considerations. This extra population might be caused by direct acceleration of the electrons by the THz field (without phonon scattering), which is possible since the THz pulse duration is comparable to the electron-phonon scattering time. A strong increase of THz absorption was also observed in InSb by high intensity THz transmission[28] and THz-pump/THz-probe[13] measurements and this effect was explained by the impact ionization process. Although Ge has a band gap and impact ionization threshold[24] more than three times larger than InSb, the field of the THz pulses with the highest energy should be enough to achieve electron energies larger than the band gap, in the absence of significant energy relaxation during the pulse. Hence, from a purely energetic point of view we cannot rule out impact ionization as the cause of a small absorption increase at intermediate energy. However, this explanation is unlikely because of two reasons. First, increased absorption from impact ionization in InSb was observed to occur after the THz pulse[13], but not during it. Second, the largest effect of impact ionization should be expected at the highest THz energies and not at intermediate energies.

We have investigated the differences of inter- and intravalley dynamics of hot free electrons in bulk semiconductors in the absence of band-to-band excitation by time- and frequency-resolved THz-pump/THz-probe measurements. We have observed unique effects in bulk Ge that we attribute to a significantly higher than thermalized population of the Γ valley of the conduction band whose mobility is higher than that of the lowest-energy valley. The behavior is simpler in Si and GaAs, whose side valleys have lower mobilities than their initially populated valleys. Although some of our quantitative results contain a large part of uncertainty, since they are based on the empirical Drude model, the qualitative features of our data are reproduced. More quantitative insight can be achieved by Monte Carlo simulations[24] taking into account the contribution of different scattering mechanisms and the full band structure which are beyond the scope of this article. Experimentally, direct observation of the intervalley and intravalley dynamics may be achieved by following pulsed THz excitation with infrared probe pulses, in order to monitor absorption or stimulated emission from individual conduction band valleys.

This work was supported in part by ONR grant no. N00014-06-1-0459.

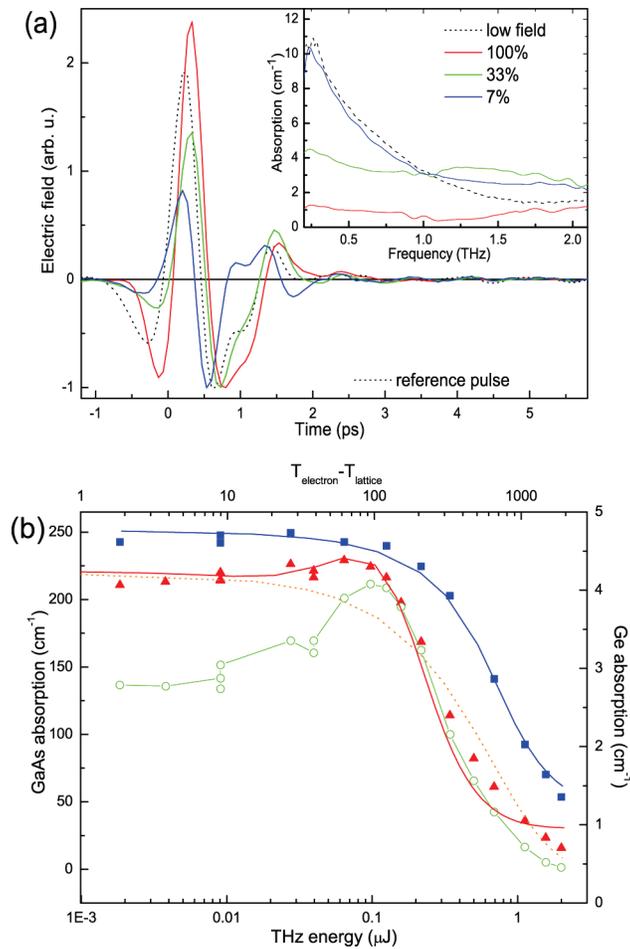

Figure 1: (color online) (a) Incident (dotted black) and transmitted THz pulse field profiles for 6 mm thick germanium at 100%, 33% and 7% of the maximum THz field strength. The temporal profiles are normalized to the second negative peaks. The inset shows the corresponding absorption spectra and the low-field (dashed) absorption spectrum measured with 9 orders of magnitude lower THz pulse energies. (b) Frequency-integrated THz absorption versus THz pulse energy for GaAs (squares) and Ge (triangles), at room temperature. Also shown is the Ge absorption integrated in the 1.0-1.3 THz range (open circles) which highlights the increase in absorption at intermediate THz field levels. The solid and dotted curves are results of model calculations. See text for details.

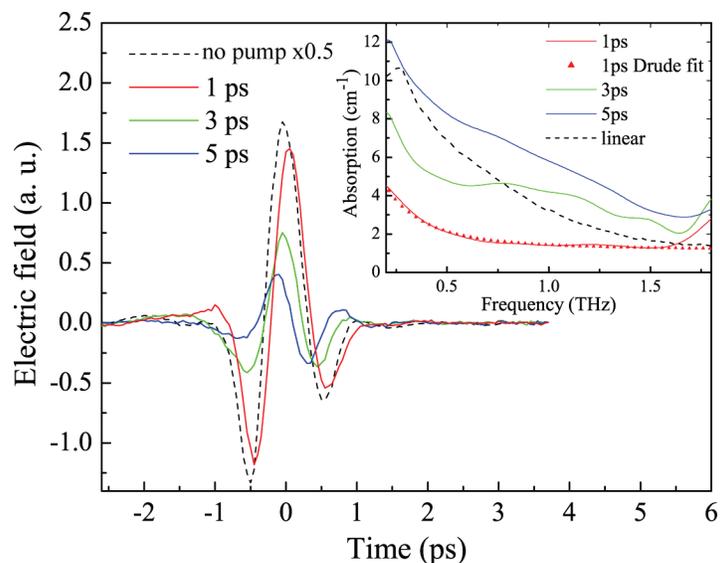

Figure 2: (color online) The incident (dotted black, scaled by a factor of 0.5) and transmitted THz probe field profiles for 6 mm thick germanium measured with the THz-pump/THz-probe setup for 1, 3, and 5 ps probe delays, respectively. The electric fields are measured in arbitrary units but the scale for the different transmitted traces is the same. The inset shows the corresponding absorption spectra, the linear absorption spectrum (dotted black), and a fit to the sum of two Drude-type components at 1 ps (red triangles).

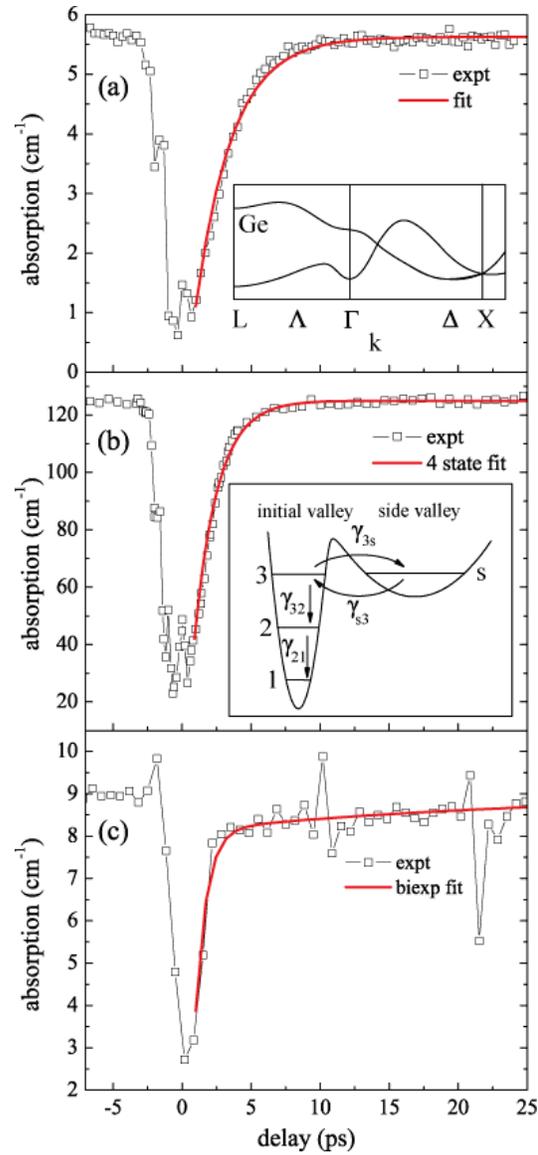

Figure 1: Spectrally averaged probe pulse absorption versus THz probe delay for (a) Ge, (b) GaAs, and (c) Si. The insets show (a) schematic illustrations of the conduction band structure for Ge and (b) the states and transfer rates of a simplified rate equation model.